\documentclass[11pt]{article}
\pdfoutput=1

\usepackage{amsbsy,amsfonts,enumerate,epsfig,graphicx,rotating,subfigure}

\newcommand{\beq}{\begin{equation}}
\newcommand{\eeq}{\end{equation}}
\newcommand{\beqa}{\begin{eqnarray}}
\newcommand{\eeqa}{\end{eqnarray}}

\begin{document}


\title{A method that reveals the multi-level ultrametric tree hidden in p-spin glass like systems}

\author{R. Baviera $^{\dagger *}$\footnote{email:roberto.baviera@gmail.com}, M.A. Virasoro$^{\dagger **}$ \footnote{email: mvirasoro@gmail.com}\\
$^{\dagger}$ Dipartimento di Fisica, {\em Sapienza} Universit\`a di Roma \\
$^*$Department of Mathematics, Politecnico di Milano\\
$^{**}$ Instituto de Ciencias, Universidad Nacional General Sarmiento\\}

\maketitle
\begin{abstract}

In the study of disordered models like spin glasses the key object of interest is the rugged energy hypersurface defined in configuration space. The statistical mechanics calculation of the Gibbs-Boltzmann Partition Function gives the information necessary to understand the equilibrium behavior of the system as a function of the temperature but is not enough if we are interested in more general aspects of the hypersurface: it does not give us, for instance, the different degrees of ruggedness at different scales. In the context of the Replica Symmetry Breaking (RSB) approach we discuss here a rather simple extension that can provide a much more detailed picture. The attractiveness of the method relies in that  it is conceptually  transparent and the additional calculations are rather straightforward. We think that this approach reveals  an ultrametric organisation with many levels in models like p-spin glasses when we include  saddle points. In this first paper we present the detailed calculations for the spherical p-spin glass model where we discover that the corresponding decreasing Parisi function $q(x)$ codes this hidden ultrametric organisation. 

\end{abstract}

\section{Introduction}

Some time ago we proposed \cite{Baviera:bmH-HhkF,Virasoro:2014uk,Baviera:1997vd} a method to introduce more control parameters (in addition to the temperature) in order to analyse the detailed ruggedness of an energy hypersurface of a disordered complex system. 

In so doing we were following a long line of research that considers systems composed of several real replicas (in addition to the virtual replicas of the Replica Trick) satisfying various constraints \cite{Parisi:1989jy, Franz:1993wp, Kurchan:1993tr, Franz:1995vd,
Monasson:1995uh}.

Our proposal here pushes the assumption of ultrametricity to new levels as we will be investigating the distribution of saddle points consistent with ultrametricity. The skeptical reader may perceive this as an unintended and unjustified consequence of past successes with the hypothesis. We hope that the pragmatic reader will instead appreciate the elegance and simplicity of the calculations and join us in hoping that this extension be proved correct. In any case the  cost of configurations that do not comply with ultrametricity can be analysed by other methods(see for instance \cite{Franz:1993wp,Kurchan:1993tr,Cavagna:1997ba,Cavagna:1997ve}). 

In pursuing this approach we were inspired by Derrida's {\it Generalised Random Energy model} \cite{Derrida:1985dx} where a multilevel ultrametric organisation built in the model can be revealed or hidden in the Gibbs-Boltzmann measure depending on the parameters of the model.   

Let us then imagine a configuration space partitioned in a hierarchical way along an ultrametric tree with $K+1$ layers labeled $(0,1...K)$ ($K$ replica symmetry levels -hereinafter RSB- levels). The leaves of the tree are the ergodic pure states and define regions in configuration space limited by the Edward-Anderson overlap parameter $q_K =q_{EA}$.   Each region is characterised by some kind of aggregate energy and by its size. These regions are then grouped in clusters limited by the overlap $q_{K-1}$ that defines a larger scale partition. These clusters are themselves grouped in superclusters in a scheme that continues along the tree. At each level we will have some kind of generalised free energy for every cluster/supercluster. The $q_k$'s satisfy $q_K> q_{K-1}\cdots> q_1> q_0$. Then the generic question at any level of the tree is how many $k$th clusters of overlap $q_k$ and generalised free energy $F$ are contained in a higher level cluster of overlap $q_{k-1}$ and another particular generalised free energy $F'$. This information is relevant,
for instance, for the relaxation dynamics of the system or for the design of an efficient search algorithm of the minima and cannot, in general, be derived from the partition function
$$Z_J=\sum_{\{{\mathcal C}\}} e^{-\beta H_J[{\mathcal C}]}$$
where there is a single parameter $(\beta)$ because the sum
aggregates information that one needs to keep apart. 
As usual the subscript $J$ reminds us that in the definition of $H_J$ there are an
infinite number of random quenched parameters and that we are dealing with an ensemble of hypersurfaces. The results take the form of
analytic expressions for average quantities including correlations among local
saddle points at variable distance in configuration space. From now on we omit the
$J$ subscript and unless explicitly stated we will be referring to quantities averaged over the $J$.

The remaining of the paper is organized as follows: in section 2, subsection 2.1, we first show how to introduce another parameter in addition to the temperature  to desaggregate the contributions coming from states with different free energies and configurational complexity so as to "see" smaller or shallower states that would not contribute to $Z$ because of their small weight and larger or deeper ones that do not contribute because of their reduced number.. At this point our method turns out equivalent to Monasson's proposal \cite{Monasson:1995uh}. The advantage of our approach becomes apparent in subsection 2.2 where a generalisation that adds up to $K$ parameters is considered.

In section 3 we present in full details the case of the p-spin spherical model for a hierarchical tree with 3 levels ($K=2$ RSB levels). The generalisation for any $K$ is straightforward. In this example we will use what is known about the solutions to the Thouless-Anderson-Palmer (TAP) equations. We will show that the corresponding ultrametric tree is revealed by the monotonic decreasing $x(q)$ function \cite{Kurchan:1993tr}. The presentation will show that in general, for any system, in a particular region of the constraint parameters (when the number of real replicas is large enough) the free energy of the system \emph{per replica} becomes equal to a saddle point solution of the unconstrained system. 

In section 4 we discuss some preliminary conclusions.

\section{The proposal: real replicas organised ultrametrically}

To simplify the presentation we divide the proposal in 2 subsections and assume $K=2$. It will become obvious how to generalise it for arbitrary $K$. 
\subsection{Revealing the states when their statistical weight is too small}

In this case we consider the partition function of a system
composed of $R$ real replicas, all of which have mutual overlap
$q$. Specifically:
\beq
Z(\beta,R,q)=\sum_{\{{\mathcal C}_a; a=1,\dots, R\}}{
e^{-\beta\sum^R_{a=1}H[{\mathcal C}_a]}
\prod_{\{a,a';a\neq a' \}} \delta({\mathcal C}_a \cdot {\mathcal
C}_{a'}-q) } 
\eeq
where ${\mathcal C}_a \cdot {\mathcal C}_{a'}$ is the overlap between the configurations ${\mathcal C}_a$ and ${\mathcal
C}_{a'}$ and both $a, a'$ run from 1 to R. At this moment $R$ is larger than 1 and integer but, as usual, after the calculations we will derive an analytic expression and treat $R$ as a real number. Now $q$ delimits roughly a region: due to the properties of the energy hypersurface if a configuration has a finite weight there will be a relatively large number of configurations nearby that will also contribute to $Z$. Ultrametricity implies that these configurations lie inside a ball with maximum overlap $q_{EA}$. If $q\geq q_{EA}$ all the R replicas will be inside that ball and if $q=q_{EA}$ the entropy per replica will be maximal. The calculation is particularly simple in this case. Therefore we choose:
\beq
{\partial \log Z(\beta,R,q)\over\partial q}=0 \Rightarrow q=q_{EA}
\eeq
We can also write
\beqa
\log Z(\beta,R)=-\beta R E_{\mathcal S} +R S_{\mathcal S}+\log {\mathcal N}_{\mathcal S\label{legendre1}}\\
F_{\mathcal S}= E_{\mathcal S} - {1\over \beta} S_{\mathcal S}=
 {1 \over \beta}{\partial \log Z(\beta,R)\over \partial R}  \label{derivs0}\\
\log {\mathcal N}_{\mathcal S}=\log Z(\beta,R)+\beta R F_{\mathcal S} 
\eeqa
where we identify $E_{\mathcal S}$ as the internal energy, $S_{\mathcal S}$ the state entropy while ${\mathcal N}_{\mathcal S}$ is
the number of states. 

These equations, equal to those in ref\cite{Monasson:1995uh}, separate the number of configurations per state and the
number of states.  They should be read in general as a definition of $F_{\mathcal S}$, the free energy per replica, to be identified with previous definitions when the dependence of $F_{\mathcal S}$ in $R$ is sufficiently small. Mathematically they reveal a multidimensional Legendre Transformation allowing a change of variables from $R$ to $F_{\mathcal S}$ and therefore the number  of states expressed as a function of the free energy follows and the Legendre transform properties imply:
\beq
\frac{\partial \log {\mathcal N_S}}{\partial F_{\mathcal S}}= \beta R \label{legendreprop}
\eeq

Parenthetically we observe that the lowest energy configuration in a state (the bottom of the corresponding valley) is determined implicitly by the
equation:
\beq
S_{\mathcal S}(E_{\rm bottom})=0
\eeq
It is then evident that 
\beq
E_{\rm bottom}\approx E_{\mathcal S}- \frac{S_{\mathcal S}(E_{\mathcal S})}{{\partial S_{\mathcal S}(E_{\mathcal S})\over \partial E_{\mathcal S}}}=F_{\mathcal S}
\eeq
So $F_{\mathcal S}$ is the linear extrapolation estimate of $E_{\rm
bottom}$ from an expansion of $S_{\mathcal S}$ around $E_{\mathcal S}$. Similar arguments will apply going up along the tree, the free energy of a cluster will be a good estimator of the lowest free energy state inside the cluster. This is relevant for the feasibility of a hierarchical search, and is true in practically any system. In the last section with the conclusions we will discuss what goes wrong in p-spin glass systems: why in this case hierarchical search, for instance, simulated annealing will not work in spite of this general property.

\subsection{Clusters of states}

We now consider $R_1$ replicas of groups of $R_2$ configurations such that the
latter are constrained to have an overlap $q_2$,  while replicas belonging to
different
groups have an overlap $q_1<q_2$.

Now the number of states defined by $q_2$ is decomposed in two components: number of states
belonging to a cluster ${\mathcal N}_{{\mathcal S}\in {\mathcal V}}$ and
number of
clusters ${\mathcal N}_{{\mathcal V}}$.

Then
\beq
Z(\beta, R_1, R_2, q_1, q_2)=\mathop{\sum}_{\{{\mathcal C}_{ab;a=1\dots R_1; b=1\dots R_2}\}}
e^{-\beta \sum_{a,b} H[{\mathcal C}_{ab}]}\Pi_{a,b\neq b'}\delta({\mathcal C}_{ab}\cdot {\mathcal
C}_{ab'}-q_2)\Pi_{a\neq a',b,b'}\delta({\mathcal C}_{ab}\cdot{\mathcal C}_{a'b'}-q_1)\ \eeq
where the indices $a, a'$ run from 1 to $R_1$ while the $b,b'$ run from 1 to $R_2$. We will again choose the $q_1,q_2$ such that
\beq
\frac{\partial \log Z(\beta, R_1, R_2, q_1, q_2)}{\partial q_1}=\frac{\partial \log Z(\beta, R_1, R_2, q_1, q_2)}{\partial q_2}=0
\eeq
Then $\log Z$ is a function of only the $R'$s and $\beta$ and as in eq(\ref{legendre1}) can be decomposed into:
\beq\log Z(\beta, R_1, R_2)=-\beta R_1R_2 E_{\mathcal S}+R_1R_2S_{\mathcal S}+R_1\log {\mathcal N}_{{\mathcal S}\in {\mathcal V}}+\log {\mathcal N}_{{\mathcal V}} \label{logzr1r2}
\eeq
This equation, as eq(\ref{legendre1}) should be read as a multidimensional Legendre transform and thus be complemented with
\beqa
\frac{\partial \log Z(\beta, R_1,R_2)}{\partial R_2}&=&R_1 \beta F_{\mathcal S} \nonumber \\
\frac{\partial \log Z(\beta, R_1,R_2)}{\partial R_1}& \equiv & \beta F_{\mathcal V}=R_2 \beta F_{\mathcal S}+\log {\mathcal N}_{{\mathcal S}\in {\mathcal V}},\label{derivs1}
\eeqa
the meaning of which is rather transparent: the first equation reads as another definition of $F_{\mathcal S}$, for a replica interacting with $R_1 R_2-1$ other replicas while the 2nd can be read as a definition of the free energy of a cluster. The equation corresponding to eq(\ref{legendreprop}) in this case is:
\beq
\frac{\partial \log{\mathcal N_V}}{\partial F_{\mathcal V}}=\beta R_1 \label{ prop2}
\eeq 

In these formulae $\log {\mathcal N}$ means $\log \overline{\mathcal N}$ rather than $\overline{\log {\mathcal N}}$. We will be using them where they are positive and extensive or at least the sum is positive and extensive. In the latter case, for instance if  
$$\log {\mathcal N}_{{\mathcal V}} >0; \log {\mathcal N}_{{\mathcal S}\in {\mathcal V}} <0; \log {\mathcal N}_{{\mathcal V}}>\log {\mathcal N}_{{\mathcal S}\in {\mathcal V}} $$ 
then we can read the $ \log {\mathcal N}_{{\mathcal S}\in {\mathcal V}}{\mathcal N}_{{\mathcal V}} $ as the total number of states with free energy $F_{\mathcal S}$ contained in clusters of a certain free energy $F_{\mathcal V}$

We observe that configurations inside a state are
weighed by the parameter $\beta$, while states inside a cluster use the
parameter $\beta R_2$ and clusters themselves $\beta R_1R_2$.

The generalisation of this approach to a generic ultrametric tree with $K$ total levels is now straightforward so we spare the reader a cumbersome notation. A warning however is worth repeating: as we have stressed we will choose all the overlaps $q_k$'s of the constraints to coincide with the saddle points of the system. This is the key to the simplicity of the calculations and implies that all relevant triangles among both real and virtual replicas satisfy ultrametricity.
 
\section{Calculations for the spherical p-spin glass}

In this section we detail the calculations for the p-spin spherical model. The Hamiltonian is \cite{Crisanti:1992wf}

\beq
H=-\sum_{i_1> i_2\dots>i_p} J_{i_1i_2\dots i_p} s_{i_1} s_{i_2}\dots
s_{i_p}
\eeq
where the $s_i$ are real variables subject to a spherical 
constraint: $\sum_{i=1,N} s_i^2=1$ and we have to calculate eq (9).

We apply the canonical Parisi trick (\cite{Crisanti:1992wf},\cite{Kurchan:1993tr})to derive the Free Energy per spin as a function of a $n R_1 R_2\times n R_1 R_2$  matrix ${\bf O}^{\alpha,\beta}$:

\begin{equation}
nF = - \frac{\beta}{4}\,\, \sum_{\alpha\beta}
({\bf O})^p
- \frac{1}{2\beta} Tr[\ln {\bf O}]
\label{uno1}
\end{equation}
 We can represent $\bf{O}$ as a matrix of matrices. The natural way to consider the ${\bf O}$ matrix is to order the columns (rows) lexicographically with 3 indices:the slowest one $a$ indexing the clusters and running from 1 to $R_1$, then $b$ indexing the states and running from 1 to $R_2$ and finally $j$ indexing the $n$ virtual replicas. This generalises the parametrisation proposed in \cite{Kurchan:1993tr}.
\beq
 {\bf O}=
 \bordermatrix{
     &   &   &   &  &  &  &  &\cr
     &O^{11,11} &O^{11,12} &\cdots&O^{11,1R_2} &O^{11,21}&\cdots & \cdots &O^{11,R_1R_2}\cr
     &O^{12,11} &O^{12,12} &\cdots&O^{12,1R_2}&O^{12,21}&\cdots&\cdots&O^{12,R_1R_2}\cr
       &  \vdots &   &   &  &  &  &  \cr
     &O^{R_1R_2,11}&O^{R_1R_2,12}&\cdots&O^{R_1R_2,1R_2}&O^{R_1R_2,21}&\cdots&\cdots&O^{R_1R_2,R_1R_2}\cr
     }
\eeq
where $O^{ab,a'b'}$s is a $n\times n$ matrix that encodes the overlap distribution of the $(a b)$ with the $(a' b')$ real replicas. The diagonal terms of these matrices satisfy the constraints: 
\beqa
\mbox{Diagonal terms of  } O^{ab,ab}&=&1 ;\;\; \forall a,b \nonumber \\
\mbox{Diagonal terms of  } O^{ab,ab'}&=&q_2;\;\; \forall a,b\neq b'\\
\mbox{Diagonal terms of  } O^{ab,a'b'}&=&q_1;\;\; \forall a\neq a',b, b' \nonumber \label{diag}
\eeqa
Clearly up to this point nothing is imposed on off-diagonal terms that relate different virtual replicas.
Then we make the natural ansatz:
\beqa
O^{ab,ab}&=&Q \;\;\;\;\;\;\; \forall a,b \nonumber \\
O^{ab,ab'}&=&P_2 \;\;\;\;\;\;\; \forall a,b\neq b'\\
O^{ab,a'b'}&=&P_1 \;\;\;\;\;\; \forall a\neq a',b\neq b' \nonumber
\label{matQP}
\eeqa
with $Q,P_1,P_2$ $ n \times n $ Parisi matrices with the same block sizes $m_1\times m_1,m_2\times m_2$ as in the standard RSB approach for virtual replicas with 2 RSB levels.. Once the limit $n\rightarrow 0$ is taken the corresponding functions $q(x), p_1(x) ,p_2(x)$ will have the usual probabilistic interpretation. For instance the probability distribution of ${\mathcal C}_{11}\cdot{\mathcal C}_{12}$ will  be encoded into $p_2(x)$. Therefore the 3 functions must be monotonous increasing for the formalism to make sense.

We now restrict our consideration to the case where 
\beqa
q(x)&=&q_2 \theta (x-m_2)+q_1 \theta (m_2-x) \theta (x-m_1) \\
p_2(x)&=&q(x)\\
p_1(x)&=&q_1\theta(x-m_1).
\eeqa
where $ \theta(\cdot) $ is the Heaviside function and $m_2 >m_1$.
There are 2 different levels of assumptions behind these ansatz, as stated in the previous section. The first one has to do with the diagonal terms of the matrices being chosen equal to saddle point of $\log Z(\beta, R_1, R_2)$. The second is that \underline{all} triplets of configurations ${\mathcal C}_{ab},{\mathcal C}_{a'b'},{\mathcal C}_{a''b''}$ obey ultrametricity. We conjecture, and we have checked it in some examples, that this 2nd statement is a consequence of the 1st one.

 It is basically this second assumption that simplifies the calculations because if all triangles are ultrametric then there must be a way to reshuffle rows and columns of ${\bf O}$ to write a new matrix ${\bf O^U}$ explicitly ultrametric as a Parisi  $n R_1 R_2 \times n R_1 R_2 $ matrix with 2 levels of Replica Symmetry Breaking. It is easy to check that this matrix will have block sizes $m_2^U \times m_2^U, m_1^U \times m_1^U$ with parameters $( m_2^U=m_2 R_2 , ,m_1^U= m_1 R_1 R_2)$. \emph{The calculation of the $\log Z(\beta, R_1, R_2)$ is then identical to the calculation for the unconstrained case just replacing the $m_i$ variables by the $m_i^U$ ones.}
 
 From eq (57) in \cite{Kurchan:1993tr} we write:
 
 \beqa
lz &\equiv &\frac{\log Z(\beta,R_1,R_2)}{R_1 R_2} =  - \frac{\beta^2}{4}
\{ q_1^p(m^U_2-m^U_1)+ q_2^p(1-m^U_2)-1 \} \nonumber \\
&+&  \frac{1}{2} \{ \frac{1}{m^U_1} \ln
(\Sigma_1) + (\frac{1}{m^U_2}-\frac{1}{m^U_1})\ln (\Sigma_2)+
(1-\frac{1}{m^U_2})\ln (\Sigma_{3}) \} \nonumber\\
\label{uno44}
\eeqa
where:
\beqa
\Sigma_1&=&1-q_2+m_2^U(q_2-q_1) + m_1^U q_1  \nonumber \\
\Sigma_2&=&1-q_2+m_2^U(q_2-q_1) \nonumber\\
\Sigma_3&=&1-q_2
\eeqa
The saddle point (SP) equations will obviously be {\em identical} to the equations for the unconstrained system with two subtle differences:
 \begin{itemize}
 \item The investigation of the fluctuations around the saddle point is complicated by the fact that some of the matrix elements are fixed by the constraints.
 \item  The saddle point values for the breaking locations apply to 
 $m_2^U=m_2 R_2  \;, m_1^U=m_1 R_1R_2$ while the inequalities necessary for the probabilistic interpretation of the solution apply to $m_2,m_1$.
  \end{itemize}
 
 \emph{Therefore {\em any} solution to the saddle point equations can be interpreted probabilistically  if:}
 
 \beq 
 \frac{m_1^{U,SP}}{R_1}\leq m_2^{U,SP} \;\; \mbox{and}\;\;\frac{m_2^{U,SP}}{R_2}\leq 1 
 \eeq
 which is always possible for sufficiently large $R_1,R_2$. When this happens the $\log Z(\beta,R_1, R_2)$ becomes equal to $R_1 R_2$ times the $\log Z$ of the unconstrained system and both $\log {\mathcal N_V}$ and $\log{\mathcal N_{S\in V}}$ are zero. In fact when $R_1 (R_2)$ grows we probe deeper clusters(states in the cluster) and eventually we hit the corresponding bottom configuration where $\log {\mathcal N_V}$ ($\log{\mathcal N_{S\in V}}$) are zero. It is interesting to notice here that contrary to what happens in the SK model, in systems like p-spin glasses the lowest state in the lowest cluster lies higher in free energy that the absolute ground state. We will discuss this farther in the last section.

\subsection{$R_1$ and/or $R_2$ below their critical values}

 But even if we consider decreasing values of  $R_1, R_2$ reaching and surpassing their critical values $R_1^c=m_1^{u,SP}/m_2^{u,SP} ; R_2^c= m_2^{u,SP}$ we can still take advantage of the solutions of the free unconstrained system.
  
In fact when $R_2$ ($R_1$) approaches $R_2^c \;(R_1^c)$ the probability of 2 states having overlap $q_2\;(q_1)$ goes to zero as can be seen from the matrices $ Q$'s and $P$'s. Physical arguments imply that the same must be true for smaller values. Therefore
\beqa
R_1\leq \frac{m_1^{u,SP}}{m_2^{u,SP}} &\Rightarrow & m_1=m_2 \Rightarrow  m_1^U=R_1 m_2^U \nonumber\\
R_2\leq m_2^{u,SP} &\Rightarrow & m_2=1 \Rightarrow  m_2^U=R_2 
\eeqa

This means that $m_1$ and/or $m_2$ hit a boundary and $\log Z(\beta,R_1,R_2)$ is not anymore stationary on them. {\bf The system behaviour is then described by the equations of the unconstrained system but with $m_1^U, m_2^U$ now as possible control parameters}. There are 4 possible scenarios:
\begin{enumerate}
\item $R_1 >R_1^c, R_2>R_2^c$. 
\item $R_1>R_1^c, R_2\leq R_2^c$ 
\item $R_1\leq R_1^c, R_2 > R_2^c$ 
\item $R_1\leq R_1^c, R_2 \leq R_2^c$ 
\end{enumerate}
In all 4 cases, defining $lz$ as in eq(21) we have: 
\beqa
  \frac{\partial lz}{\partial q_1}&=&\frac{\partial lz}{\partial q_2}=0 \;\;\;\mbox{Saddle Point eqs in } q_1,q_2\\ \nonumber
\frac{\partial lz}{\partial 1/R_2}&=&\log {\mathcal N_{S\in V}} \\
\frac{\partial lz}{\partial 1/(R_1R_2)}&=&\log {\mathcal N_V}\nonumber
\label{derivs2}
\eeqa
In case 1 in addition we have 2 more SP equations:
\beqa
\frac{\partial lz}{\partial m_1^U}= \frac{\partial lz}{\partial m_2^U}=0\\ \nonumber 
\Rightarrow \log {\mathcal N_{S\in V}}=\log {\mathcal N_V}=0
\eeqa
that detect the lowest free energy state in the lowest free energy cluster.

In case 2 we have one more SP equation:
\beq
\frac{\partial lz}{\partial m_1^U}=0 \Rightarrow \log {\mathcal N_V}=0\\ 
\eeq
and a new control parameter $m_2^U=R_2$  that allows us to probe states in the lowest free energy cluster
with varying free energies above the minimum one.

In case 3 again we have one additional SP equation
\beqa
\frac{\partial lz}{\partial m_2^U}+R_1\frac{\partial lz}{\partial m_1^U}=0 \nonumber \\
\Rightarrow   \log {\mathcal N_ {S \in  V}}^{R_1} {\mathcal N_ V}=0
\eeqa
and $R_1$  as a control parameter to explore lowest free energy states in higher free energy clusters.

In case 4, finally, there are no additional SP equations and instead we have 2 control parameters $m_1^U=R_1R_2$ and $m_2^U=R_2$ that explore higher free energy states lying in higher free energy clusters.
 
\section{The Saddle Point equations and their interpretation in terms of $\log {\mathcal N_{S\in V}}$ and $\log {\mathcal N_V}$.}

In principle one should first solve $q_1,q_2$ in terms of $m_1^U, m_2^U$ and then use eq(\ref{logzr1r2},\ref{derivs1},\ref{derivs2}) to derive expression for other functions in terms of $m_1^U,m_2^U$. But as the derivatives of $\log Z(\beta,R_1,R_2)$ with respect to the $q$'s are zero it is simpler to take the first derivatives with respect to the $m$'s at constant $q$'s and afterwards use the SP equations in $q$ to choose any pair of independent variables to express our results. 
We borrow from section 4 of ref \cite{Kurchan:1993tr} the saddle point equations for K=2 of the spherical p-spin glass and a set of variables that are convenient because their SP values are independent of $\beta$

\begin{enumerate}
\item Variables:
\beq
w_1=\frac{q_1}{q_2} ;\;\;\;y_1=\frac{\Sigma_2}{\Sigma_1};\;\;\;y_2=\frac{\Sigma_3}{\Sigma_2} ;\;\;\ Y= {p\beta^2\over 2} q_2^{p-2} (1-q_2)^2
\eeq
\item  SP equations in $q_1,q_2$  
\beq
 {p\beta^2\over 2} (q_s^{p-1}-q_{s-1}^{p-1})=\frac{q_s-q_{s-1}}{\left[
\Sigma_{s+1} \Sigma_{s} \right]}\;\;\;\;\; s=1, 2 \;\;\;  q_0=0\nonumber \\
\eeq
that imply one equation among the new variables:
\beq
w_1^{1-p}-1=\frac{w_1^{-1}-1}{y_1 y_2} \label{w1}
\eeq
and another equation $\beta$ dependent;
\beq
{p\beta^2\over 2} q_1^{p-2}=\frac{1}{\Sigma_{2} \Sigma_{1}}\;\;\Rightarrow\;\; y_1= {p\beta^2\over 2} q_1^{p-2} \Sigma_2^2 \label{sp1}
\eeq

that implies a 2nd equation for $Y$
\beq Y=y_1 w_1^{2-p} y_2^2  \label{w1y2}
\eeq
\item Variations with respect to $m_1^U, m_2^U$
\beqa
{-(m_1^U)^2}\frac{\partial lz}{\partial m_1^U}&=&\log {\mathcal N_V}=\\  \nonumber
&=& -\frac{1}{2}(1-y_1+\log y_1+\frac{(1-y_1)^2}{py_1})\\
{- (m_2^U)^2}\frac{\partial lz}{\partial m_2^U}&=&\log {\mathcal N_{S\in V}}=\\ \nonumber
&=&-\frac{1}{2}(1-y_2+\log y_2+\frac{(1-y_2)^2}{py_2} \frac{(1-p(1-w_1)w_1^{p-1}-w_1^p)}{(1-w_1)(1-w_1^{p-1})})
\eeqa
\end{enumerate}

We remark the equality of the expression for the $\log {\mathcal N_V}$ and the corresponding one for the $\log {\mathcal N_S}$ derived in  \cite{Crisanti:1995ux}, both of them independent of the temperature when expressed in terms of the corresponding variables:
$ y_1$ and $Y=(p\beta^2/2) q_2^{p-1} (1-q_2)^2 $ respectively. This fact is evidence that the multiplicity of states and clusters is connected to the multiplicity of solutions at $T=0$, a fact that could be checked studying the TAP equations for the clusters derived from the Cavity Method. From \cite{Crisanti:1995ux,Kurchan:1993tr} we know that in the range of variations of Y the $\log {\mathcal N_S}$ is 0 when $Y=0.354993$ for p=3 while at  $Y=1/(p-1)$  all states become unstable. The lower bound for $y_1$ is obviously the same but the higher one would depend on a stability analysis that has not been done, we are not distinguishing between stable or unstable clusters.\footnote{In a work in progress by G. Parisi, F. Ricci-Tersenghi and M.A.V. we have shown that the Plefka stability criterion implies $y_1 \leq 1/(p-1)$}

\section{Conclusions and future program}

Using eqs(\ref{w1}) and (\ref{w1y2}) we express $\log {\mathcal N_{S\in V}}$ in terms of $Y$ and $y_1$. In Fig1(a) we show the contour plot of this function. We observe that the boundary of the region where there are an exponentially large number of states inside clusters has a positive derivative. This means that if we choose two clusters with different $y_1$ the one with the largest value contains states with larger $Y$, a non surprising result if as expected $y_1$ and $Y$ are related to the free energies of clusters and states. The same result is expected in a model like the Sherrington-Kirkpatrick model and should in general be true. In a cluster with lower free energy we expect to find the lowest free energy state because as proved by definition the cluster free energy is an estimate of the free energy of the lowest states inside it.

\begin{figure} 
\subfigure[]{ 
\resizebox*{8cm}{!}{\includegraphics{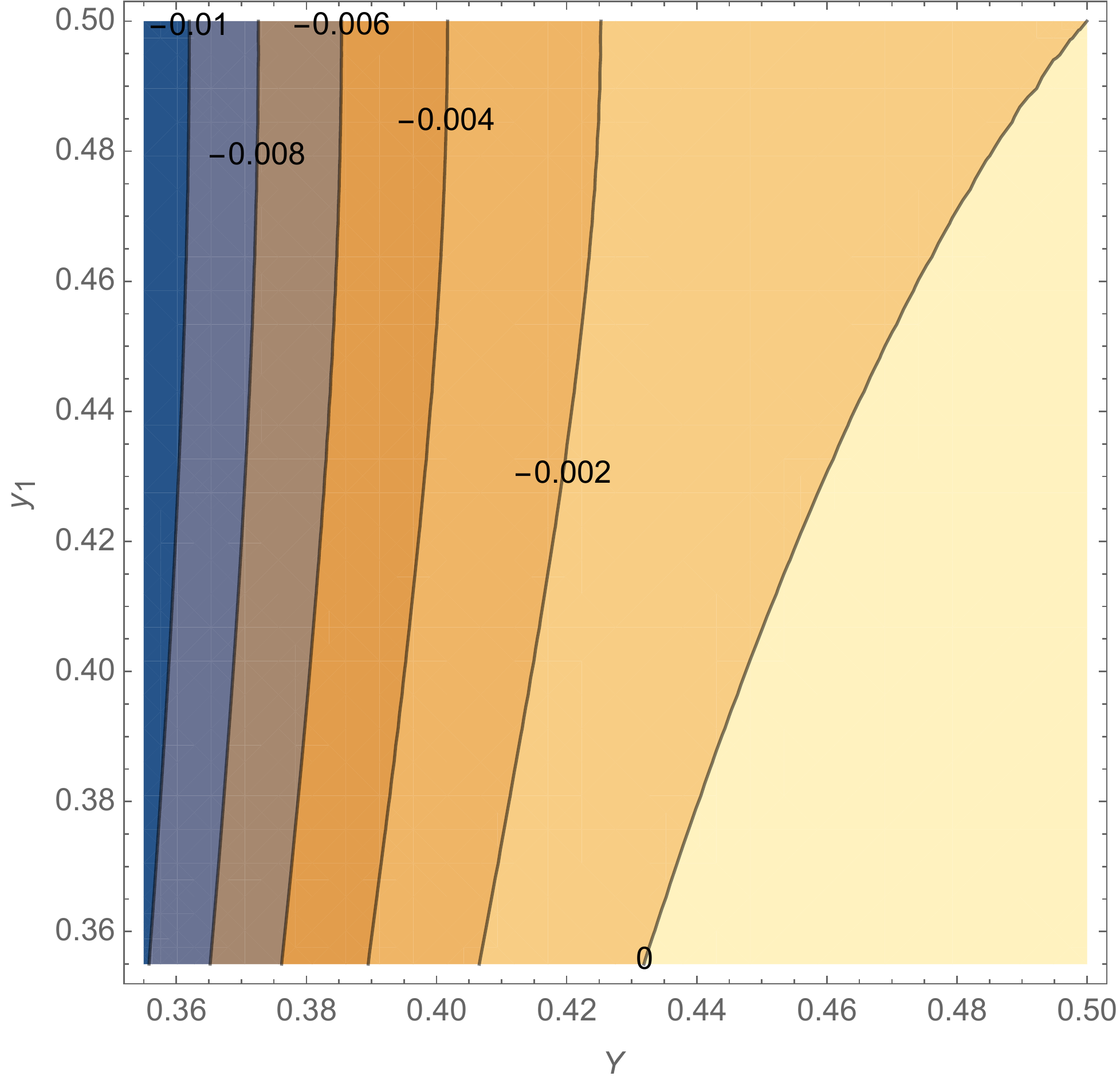}}}%
\subfigure[]{ 
\resizebox*{8cm}{!}{\includegraphics{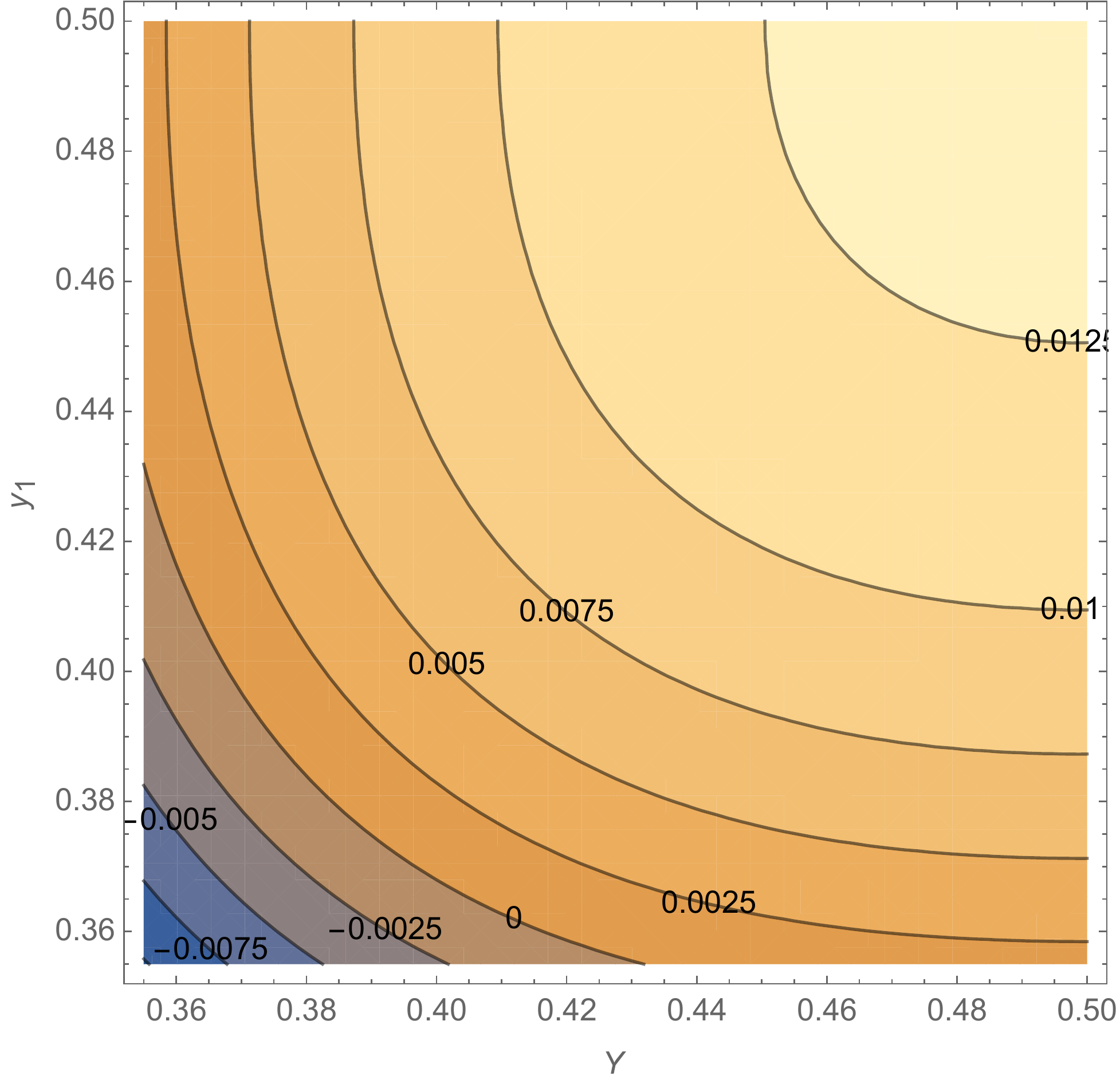}}}%
\label{sample-figure}
\caption{(a) Contour-plot of  $\log {\mathcal N_{S\in V}}$ as a function of $y_1$ and $Y$. (b) Contour-plot of $\log {\mathcal N_{S\in V} N_S}$ as a function of $y_1$ and $Y$.}
\end{figure}

On the other hand we observe that the lowest value of $Y$ in Fig1(a) lies around 0.43 a value much larger than the ground state value 0.354993 found in \cite{Kurchan:1993tr}. So the question: where is the ground state? This is explained in Fig1(b) where we draw the contour plot of the $\log {\mathcal N_V N_{S\in V}}$, i.e. the log of the total number of states labeled by $Y$ contained in all clusters labeled by $y_1$ and we witness a dramatic change of the derivative of the contour line 0. What we are finding is that the number of clusters grows exponentially with $y_1$ at such a rate that even if the probability of finding a state with lower $Y$ in a typical cluster goes exponentially to  zero there remains a fraction of clusters that contain such states. This is exactly the picture suggested by the GREM and we believe it is common to all systems where one finds a decreasing $x(q)$. In fact in the linear approximation around the lowest state in the lowest cluster:
\beqa
\log {\mathcal N_V}(F_{\mathcal V})\cong \beta x(q_1) (F_{\mathcal V}-F_{\mathcal V,0})\nonumber \\
\log {\mathcal N_{S\in V}}(F_{\mathcal S})\cong \beta x(q_2)(F_{\mathcal S}-F_{\mathcal V})
\eeqa

so the contour line 0 of Fig1(b) 
\beq
\log {\mathcal N_V N_{S\in V}}=0 \Rightarrow x(q_2) F_{\mathcal S}+(x(q_1)-x(q_2))F_{\mathcal V}=constant.
\eeq
Therefore $x(q_1)>x(q_2)$ for $q_1<q_2$ is telling us that an increase in $F_{\mathcal V}$ allows a decrease in $F_{\mathcal S}$.

The calculation for the spherical p-spin determines the value of $y_1$ (in Fig1(b) for $p=3$ around 0.41) that labels the lowest free energy clusters that contain the ground state. Unfortunately, and contrary to what we argue in a previous paper \cite{Virasoro:2014uk} it is not easy to use this additional information to improve on a possible hierarchical search of the ground state. At that value of $y_1$ the number of clusters is exponentially large even if with a smaller exponent and we have to visit all of them since only a very small fraction (in the limit just one) will contain the ground state. Furthermore there is no signal at that level of the search that reveals which are the good clusters. If the search inside a cluster could be done using gradient descent there would be some advantage but we do not think that even this happens.

There are many things that remain to be done in the follow-up to this work. The most immediate is the analysis of the stability of the solutions of the constrained system. Another one is to apply the method to mixtures of different p's spin glasses. There is also the idea of exploring alternative $q(x)$ for these same systems. Further work on these aspects is the subject of present work (G. Parisi, F. Ricci-Tersenghi, M.A.V., in preparation)

\begin{center}
Acknowledgements
\end{center}
One of us (MAV) would like to acknowledge relevant conversations with Giorgio Parisi and Federico Ricci-Tersenghi. Most of the results in this paper are included in Baviera's thesis \cite{Baviera:bmH-HhkF} almost 20 years old. The impulse to publish it originates basically from Parisi's interest and encouragement. 
 
\newpage

\bibliographystyle{unsrt}
\bibliography{ultrampush}

\end{document}